# COMPUTATIONAL STUDY OF THE EFFECTIVE THREE-ION INTERACTION POTENTIALS IN LIQUID METALS WITH HIGH DENSITY OF ELECTRON GAS

E. V. Vasiliu[*]

A.S. Popov Odessa National Academy of Telecommunications,
1 Kusnechnaia St., Odessa, 65021, Ukraine

Based on the many-body theory of metals in the third order of the perturbation expansion in electron-ion interaction pseudopotential, the potentials of pair and three-ion interactions are calculated in liquid lead, aluminium and beryllium at their melting temperatures. The reducible and irreducible three-ion interactions have an attractive nature on distances approximately equal to an average distance between ions in metals. It results in the shortening of average interatomic distance in an equilibrium state of metal. The potential landscapes created by a pair of fixed ions relative to the third ion are constructed. It is shown that with increasing of an electronic density the contribution as reducible, that and irreducible three-ion interaction is increased. It is shown also that the influence of reducible three-ion interaction on a potential landscape in a cluster of three ions is considerably larger than influence of irreducible three-ion interaction.

## 1. Introduction

The study of a nature of multi-ion interactions, their influence on different physical properties of metals is now the urgent problem of condensed matter physics, but very complicated and insufficiently investigated. A main task for the solution of this problem as a whole is a calculation of multi-ion interaction potentials. A few methods of calculation of effective multi-ion interactions in metals now exist. First, this is a microscopic many-body theory of metals [1], which is grounded on perturbation theory in electron-ion interaction pseudopotential. Second, the embedded atom method [2,3], and the "glue potential" method [4-6], where the multi-parameter potential functions fitted to experimental measured physical properties of metals are used. These two last methods are widely used, for example, in molecular dynamic simulations, but in fact are semiempirical methods. Gurskii and Krawczyk [7-9] proposed recently in a general view one new microscopic approach, which is grounded on density functional theory.

---

[*] E-mail: vasiliu@ukr.net

Nowadays the main microscopic method of multi-ion interaction study is the many-body theory of metals. Within the framework of this theory, some evaluations of equilibrium [1,10] and kinetic [11] properties of simple metals are already carried out. These evaluations take into consideration the contributions of the higher orders of a perturbation theory in powers of pseudopotential, which are multi-ion interactions. However, the majority of these evaluations are carried out without considering multi-ion potentials. It means that calculations of energy are carried out in reciprocating space. This is a conventional method for a crystalline state of metals, where generally it is possible does not consider interaction potentials, and usage of symmetry of a crystal allows considerably simplifying evaluations [1,10]. For homogeneous liquid state, the evaluations also can be carried out in reciprocating space [11]. However, in this case remains open a question about relative quantity of the so-called reducible and irreducible contributions in multi-ion potentials, and a question about their influence on the short-range order in a liquid. Therefore, in amorphous, liquid and inhomogeneous metals the calculations in configuration space are preferable.

Within the framework of the many-body theory of metals, it is possible correctly to separate irreducible and reducible contributions into multi-ion potentials. The last arise out from the terms of certain order of a perturbation theory in pseudopotential, when coordinates of two or more ions coincide. Hasegawa for the first time has obtained the corresponding formulas for the three-ion interaction [12]. However, the numeral computations were done only for equilateral configurations of three ions in liquid sodium and potassium. Moriarty has carried out computations of three-ion potentials for isosceles ion configurations in some transition metals [13,14]. However, systematic analysis of the three-ion potentials within the framework of same approaches for a row of metals with different valency and density of electron gas before was not carried out.

It is known that the contributions of the third order perturbation theory most essentials for polyvalent simple metals with high density of electron gas [1,10]. Therefore, in the present paper for the analysis of three-ion interactions, we chose quadrivalent lead, trivalent aluminium and divalent beryllium. The last has most density of electron gas from all simple metals. The purposes of this paper are the followings: a computation of reducible and irreducible three-ion interactions in these metals at their melting temperature; a determination of relative contribution of the three-ion interactions into full interaction potential of three ions; a clarification of the value dependence of these interactions from density of electronic subsystem of metal.

## 2. Many-body theory of metals

In the context of many-body theory of non-transition metals, the energy of a metal is being calculated with using the adiabatic approximation for the electron-ion system. The energy of electronic subsystem $E_e$, if ion positions are fixed, is written as a series in terms of powers of electron-ion pseudopotential [1]:

$$E_e = E_e^{(0)} + E_e^{(1)} + \sum_{n \geq 2} E_e^{(n)}. \tag{1}$$

Here $E_e^{(0)}$ is the energy of homogeneous electron gas, $E_e^{(1)}$ is the contribution of the first order due to undotted of ions, and a sum $\sum_{n \geq 2} E_e^{(n)}$ is the energy of band structure. Into (1) one can be separated the contributions independent of ion positions, dependent on the locations of separate ions, ion pairs, triplets, etc. Then [1]:

$$E_e = \varphi_0 + \sum_n \varphi_1(\vec{R}_n) + \frac{1}{2!}\sum_{n \neq m}\varphi_2(\vec{R}_n, \vec{R}_m) + \frac{1}{3!}\sum_{n \neq m \neq k}\varphi_3(\vec{R}_n, \vec{R}_m, \vec{R}_k) + \ldots \tag{2}$$

Each term of the series (2) describes indirect interactions of ion groups through the surrounding electron gas. Using (1) $\varphi_2$, $\varphi_3$, etc. can be represented as a series in terms of powers of pseudopotential:

$$\varphi_2(\vec{R}_1, \vec{R}_2) = \sum_{i=2}^{\infty} \Phi_2^{(i)}(\vec{R}_1, \vec{R}_2), \tag{3}$$

$$\varphi_3(\vec{R}_1, \vec{R}_2, \vec{R}_3) = \sum_{i=3}^{\infty} \Phi_3^{(i)}(\vec{R}_1, \vec{R}_2, \vec{R}_3), \tag{4}$$

where $\Phi_n^{(k)}(\vec{R}_1, \ldots, \vec{R}_n)$ represents the indirect interaction of the *n* ions through electron gas in the *k*-order perturbation theory in electron-ion interaction. So, pair interaction in the third order is the sum of a direct Coulomb repulsion of ions (we neglect by the overlapping of ion shells), indirect interaction in the second order $\Phi_2^{(2)}$, and indirect interaction in the third order $\Phi_2^{(3)}$ (the latter is called as reducible three-ion interaction):

$$\varphi_2^*(R) = (ze)^2/R + \Phi_2^{(2)}(R) + \Phi_2^{(3)}(R), \tag{5}$$

where $z$ is the ion charge.

For simple metals good approximation is a local form of electron-ion pseudopotential. Then

$$\Phi_2^{(2)}(R) = \frac{1}{\pi^2} \int_0^\infty dq\, q^2 \Gamma^{(2)}(q) |W(q)|^2 \frac{\sin(qR)}{qR}, \qquad (6)$$

$$\Phi_2^{(3)}(R) = \frac{3}{4\pi^4} \int_0^\infty dq_1 q_1^2 \int_0^{q_1} dq_2 q_2^2 \int_{-1}^1 dx\, W(q_1) W(q_2) W(q_3) \times \\ \times \Gamma^{(3)}(q_1, q_2, q_3) \left( \frac{\sin(q_1 R)}{q_1 R} + \frac{\sin(q_2 R)}{q_2 R} \right) \qquad (7)$$

Here $\Gamma^{(2)}(q)$ and $\Gamma^{(3)}(q_1, q_2, q_3)$ are the sums of two- and three-pole diagrams respectively; $W(q)$ is a form-factor of local pseudopotential; $q_3 = (q_1^2 + q_2^2 + 2q_1 q_2 x)^{1/2}$.

A first term in the series (4) is irreducible three-ion interaction in the third order of perturbation theory. It describes by the expression [12]:

$$\Phi_3^{(3)}(R_{12}, R_{23}, R_{13}) = \frac{3}{2\pi^4} \int_0^\infty dq_1 q_1^2 \int_0^\infty dq_2 q_2^2 \int_{-1}^1 dx\, W(q_1) W(q_2) W(q_3) \times \\ \times \Gamma^{(3)}(q_1, q_2, q_3) \int_0^1 dy \cos\left\{ y \left( q_1 R_{12} \frac{R_{12}^2 + R_{23}^2 - R_{13}^2}{2 R_{12} R_{23}} + q_2 R_{23} x \right) \right\} \times \\ \times J_0 \left\{ q_1 R_{12} (1-y^2)^{1/2} \left( 1 - \frac{(R_{12}^2 + R_{23}^2 - R_{13}^2)^2}{4 R_{12}^2 R_{23}^2} \right)^{1/2} \right\} \times \\ \times J_0 \left( q_2 R_{23} (1-y^2)^{1/2} (1-x^2)^{1/2} \right), \qquad (8)$$

where $J_0(x)$ is the Bessel function of the zero-order; $R_{12}$, $R_{23}$ and $R_{13}$ are the distances between the vertices of a triangle formed by the ions.

### 3. Computations Results

The pair and three-ion interaction potentials were calculated at melting point of lead, aluminium and beryllium. The correspondence values of Wigner-Seitz radiuses are $(r_s)_{Pb} = 2.3560$, $(r_s)_{Al} = 2.1677$ and $(r_s)_{Be} = 1.9185$ (all the values in atomic units). For all the metals we use local two-parameter Animalu-Heine pseudopotential with the form-factor

$$W(q) = -\frac{4\pi e^2 z}{q^2 \Omega_0} [(1+U) \cos q R_0 - U \sin q R_0 / q R_0] \exp(-0{,}03(q/2k_F)^4). \qquad (9)$$

A permittivity function in the Vashishta-Singwi form was also used [15]. It provides good description of exchange-correlation effects in an electron gas at small values of $r_s$. The parameters of the pseudopotential (9) for lead and aluminium are taken from [10], where they were fitted in the fourth order perturbation theory on requirements $p = 0$ and $C_{44} = C_{44\,(experimental)}$ ($p$ is the pressure, $C_{44}$ is the shear modulus). For beryllium the suitable data in the literature are absent, it is known only the position of the first zero of pseudopotential form-factor [16]. The aim of this paper was not examination of the three-ion potential dependence from a choice of the pseudopotential form. Therefore, for beryllium we did not carried out the precise parameter adjustment procedure. The parameter $R_0$ was taken equal to ionic radius, and the second parameter $U$ was fitted with use of condition $W(q_0) = 0$.

Figure 1 shows computed potential $\varphi_2^*(R)$ (5) and its components. For all three metals reducible three-ion interaction $\Phi_2^{(3)}(R)$ has attractive nature on short distances between two ions; therefore, the pair interatomic potential is strongly renormalizes. In the first place, the first minimum position shifts towards shorter distances. This effect is almost identical for all three metals. Secondly, a minimum depth is largely increased. This effect is already unequal: it least for lead, slightly more for aluminium and most for beryllium. In beryllium at taking into account only the second order of the perturbation theory, a depth of the first minimum of pair potential is smaller than a depth of the second minimum. Taking into account reducible three-ion interaction $\Phi_2^{(3)}(R)$ leads up to increase of depth of the first minimum almost five times. In result, the potential $\varphi_2^*(R)$ in beryllium takes the form typical for simple metals (see figure 1(c)).

In a row of metals: lead, aluminium, beryllium, the values of $r_s$ decrease while the contribution of $\Phi_2^{(3)}$ increases. Consequently, one can conclude that importance of $\Phi_2^{(3)}$ consideration increases with decrease of $r_s$. We note that this conclusion is in accordance with computations of pair interproton interaction in metallic hydrogen [17]. In this system at $r_s = 1.65$ pair potentials in the second order have not the minimum at all and only taking into account of $\Phi_2^{(3)}$ leads up to its formation.

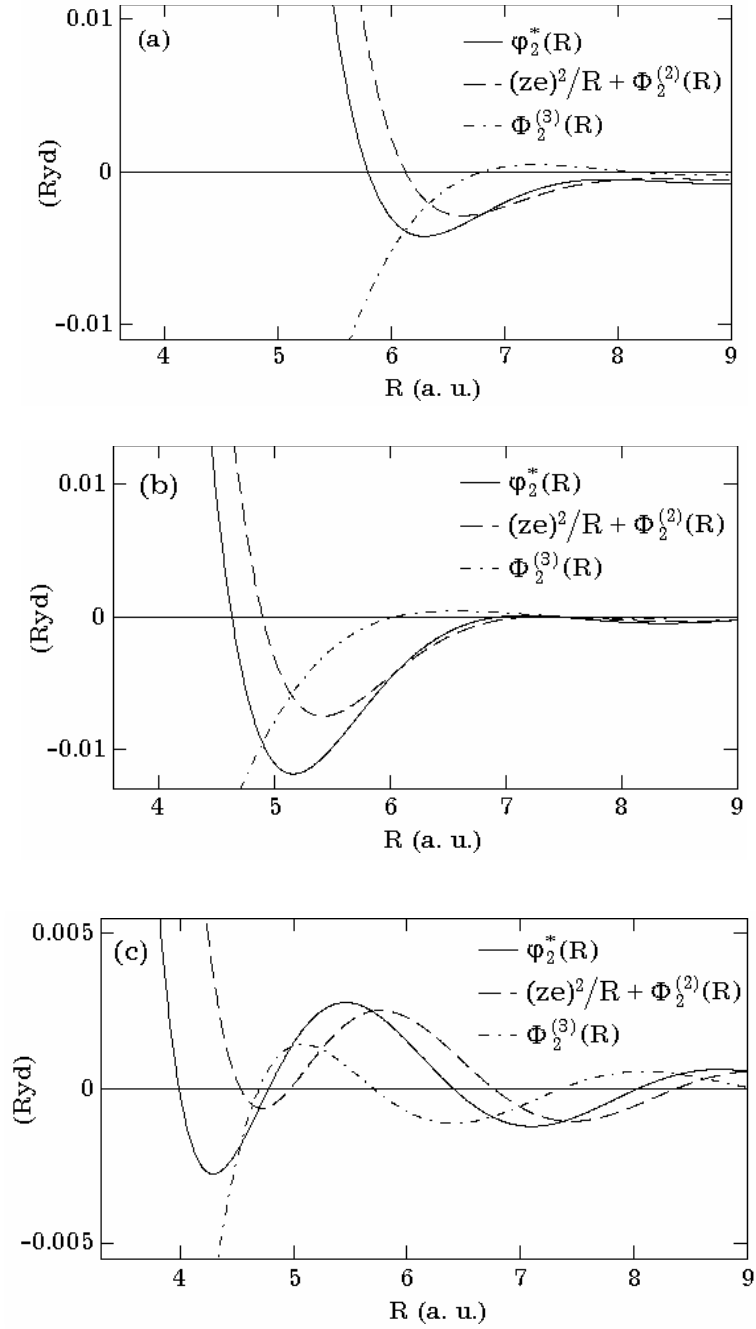

**Figure 1.** Pair interaction potential and its components corresponding to (a) lead, (b) aluminium and (c) beryllium.

The irreducible three-ion interaction potential $\Phi_3^{(3)}(R_{12}, R_{23}, R_{13})$ is the three-dimensional function and can be represented either as the table or graphically as the "sections": surfaces or curves for certain configurations of ions. So, for example, Moriarty has presented the results of three-ion potential calculations in transition metals for isosceles ion configurations as function of a vertex angle of an isosceles triangle [13,14]. For the application purposes the $\Phi_3^{(3)}$ potential should be tabulated or should be approximated by simple analytic function, however, for qualitative consideration preferably to present it as a potential landscape for a cluster of three ions (the table of $\Phi_3^{(3)}$ values for any ion configurations and its analytic approximation will be given elsewhere). We consider that such form of result representation is visual and informative, as allows clearing up the behaviour of the $\Phi_3^{(3)}$ potential and its influence on a potential landscape in a cluster.

Figure 2 shows the irreducible three-ion interaction potential for the special case where three ions form a regular triangle. The dotted line in this figure corresponds to liquid sodium at melting point. It was computed for comparing (the data for computation was taken from [12]). As well as reducible three-ion interaction potential $\Phi_2^{(3)}(R)$, the potential $\Phi_3^{(3)}(R,R,R)$ has attractive nature on short distances between ions and oscillates on large ones.

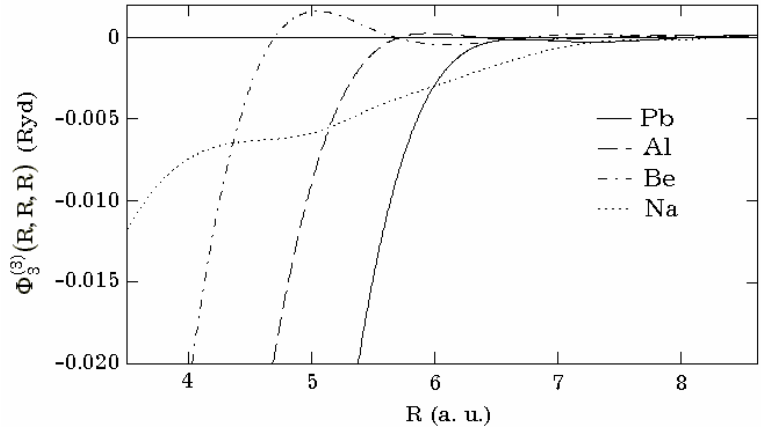

**Figure 2.** Three-ion interaction potential for a regular triangle of ions.

In figures 3 and 4, two ions are located on an ordinate axis on fixed distance one from other and the third ion is sited in the *XY*-plane. The potential landscapes in all these figures are shown only in one coordinate quarter, as they are symmetrical concerning axes *X* and *Y*. The distance between two fixed ions for each metal is selected accordingly to position of the first minimum of pair potential $\varphi_2^*(R)$. This is 6.3 a. u. for lead, 5.16 a. u. for aluminium and 4.3 a. u. for beryllium (see figure 1). We note that these positions of the first minimum of $\varphi_2^*(R)$ are in good accordance with the corresponding experimental values of most probably distances between neighbouring atoms.

The potential of irreducible three-ion interaction $\Phi_3^{(3)}$ is similar for all three metals (see figures 2-4). It has sufficiently deep potential well (in beryllium it has even two potential wells, which are separated by a saddle point). On large distances, the potential $\Phi_3^{(3)}$ has the damping oscillations. These oscillations are determined by Fridel's oscillations of electronic density. A distance from a bottom location of a potential well up to fixed ions approximately equal to an average interatomic distance for each of metals.

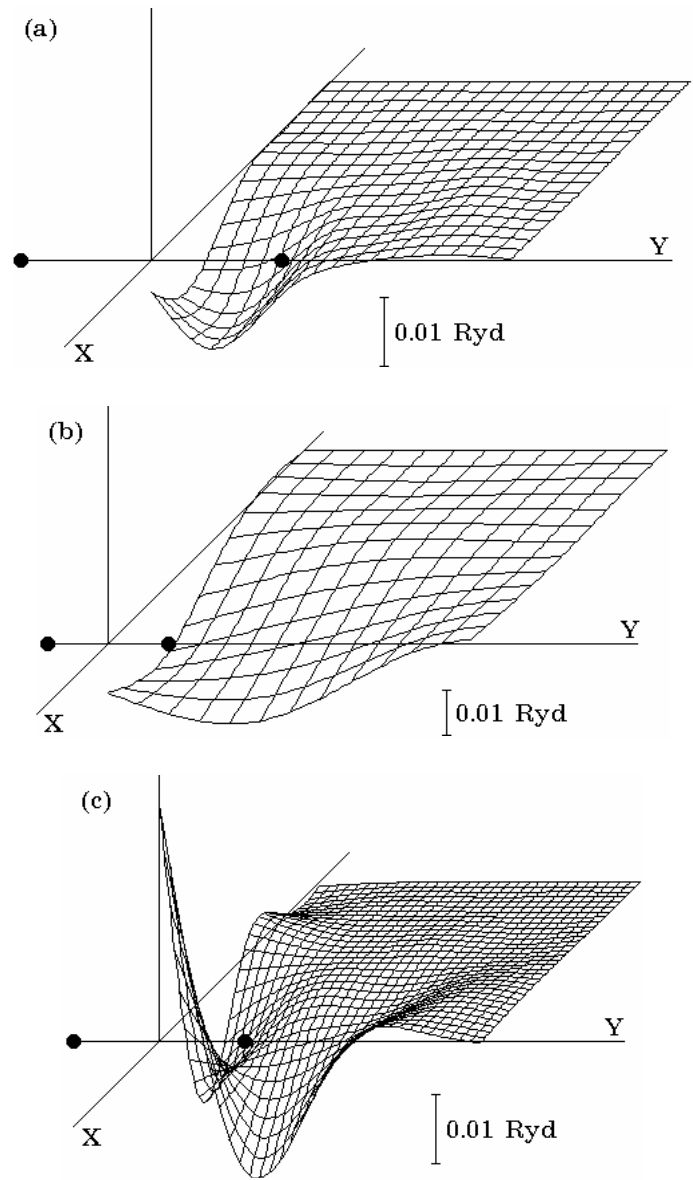

**Figure 3.** Irreducible three-ion interaction potential corresponding to (a) lead, (b) aluminium and (c) beryllium.

Figure 4 shows the potential landscape, which is created ions pair and taking into account pair- and three-ion interactions. We do not show in this figure a potential landscape for lead, as here the potential $\Phi_3^{(3)}$ does not influence almost a landscape and the last cannot be drawn in reasonable scale. The drawing has been executed on radial directions from origin of coordinates. A solid line corresponds to pair interaction in the third order taken into account that describes by the function $\varphi_2^*(R_{23}) + \varphi_2^*(R_{13})$. A dotted line corresponds to pair and irreducible three-ion interaction taken into account that describes by the function $\varphi_2^*(R_{23}) + \varphi_2^*(R_{13}) + \Phi_3^{(3)}(R_{min}, R_{23}, R_{13})$. In these formulas the indexes 1 and 2 corresponds to the fixed ions and the index 3 corresponds to the free ion. The $R_{min}$ are the positions of $\varphi_2^*(R)$ first minimum for each of metals (see above).

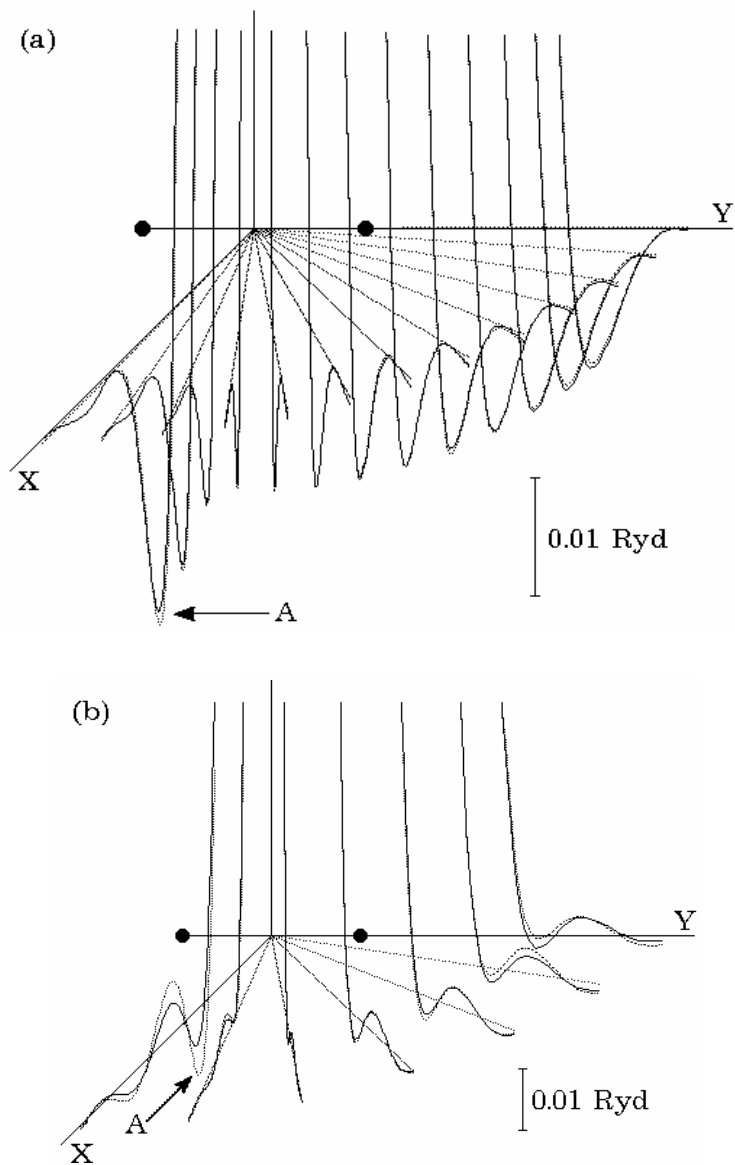

**Figure 4.** Potential landscape created by two fixed ions for the third ion corresponding to (a) aluminium and (b) beryllium.

As noted above the potential $\Phi_3^{(3)}$ does not influence almost a potential landscape in lead, it only increases insignificant (~1%) a minimum depth on some directions. In aluminium (see figure 4(a)), an effect of irreducible three-ion interaction is something more, but is also small: on ~4% deepens the minimum disposed on *X*-axis (minimum *A*). In beryllium (see figure 4(b)) the contribution of $\Phi_3^{(3)}$ is rather significant that especially noticeably in a direction of *X*-axis. In this direction, the $\Phi_3^{(3)}$ potential taking into account result in increase the depth of the first minimum approximately twice (minimum *A*) and some increase the depth of the second minimum. Thus in beryllium the effect of irreducible three-ion interaction is spread even to the second coordination shell.

## 4. Discussion of results

There is an apparent tendency in a row of simple liquid metals with various density of electron gas: with increasing of density (decreasing of $r_s$) the contribution both reducible and irreducible three-ion interactions increases.

Both reducible and irreducible three-ion interactions have an attractive nature on distances approximately equal to the equilibrium distance between ions. We note that the same nature has also indirect pair interatomic interaction in the second order $\Phi_2^{(2)}$. Thus, the attractive nature of all indirect interactions (in the second and in the third order) leads up to the shortening of average interatomic distance in an equilibrium state of metal.

It is well known that the nature of pair interatomic interaction in the second order is identical for all of simple metals [18]. Our computations and the computations for liquid sodium and potassium [12] allow drawing a conclusion that the three-ion interactions also are similar in all simple metals.

For all considered here metals the influence of reducible three-ion interaction $\Phi_2^{(3)}$ on a potential landscape in a cluster of three ions is considerably larger than the influence of irreducible interaction $\Phi_3^{(3)}$. This fact in some measure can be a ground for approach, when at calculations of the physical properties of metals the $\Phi_3^{(3)}$ is taken into account as perturbation at the accurate consideration of $\Phi_2^{(3)}$ [12], or the $\Phi_3^{(3)}$ contribution neglect and take into account only $\Phi_2^{(3)}$ [19], especially for metals with low density of electron gas. However, in recent years the properties of materials at high pressure are actively explored theoretically and experimentally. We believe that for metals at increasing pressure the role of indirect multi-ion interactions should increase essentially.